\documentstyle[11pt]{article}

\def\d{\delta}

\def\be{\begin{equation}}
\def\ee{\end{equation}}
     
\renewcommand{\thefootnote}{\fnsymbol{footnote}}

\begin{document}
\begin{flushright}
BRX TH 439\\
\end{flushright}
\setcounter{footnote}{0}

\begin{center}{\Large\bf Mapping Hawking into Unruh Thermal Properties}

S. Deser and Orit Levin\footnote{
deser@binah.cc.brandeis.edu, olevin@binah.cc.brandeis.edu}

Physics Department, Brandeis University, Waltham, Massachusetts
02454-9110, USA
\end{center}

\renewcommand{\thefootnote}{\arabic{footnote}}
\setcounter{footnote}{0}

\begin{quotation}
{\bf Abstract}: By globally embedding curved spaces into 
higher dimensional flat ones, we show that Hawking thermal properties
map into their Unruh equivalents: The relevant curved
space detectors become Rindler ones, whose temperature and entropy reproduce
the originals.  Specific illustrations
include Schwarzschild, Schwarzschild-(anti)deSitter, Reissner-Nordstrom and
BTZ spaces.     
\end{quotation}

\noindent{\bf 1.Introduction}

It is well-understood that, for both Hawking and Unruh
effects, temperature emerges from information
loss associated with real and accelerated-observer
horizons, respectively.  Given that any D-dimensional
geometry has  a higher-dimensional 
global embedding  Minkowskian (possibly with more than one timelike coordinate)
spacetime (GEMS)  \cite{goenner},
it is  natural to ask whether these mappings can unify
the two effects, by associating the relevant detectors
of the curved spaces and their horizons with (constant acceleration) Rindler
detectors and their horizons.
Confirmation of these ideas was recently given in an
analysis of deSitter (dS) \cite{n} and anti-deSitter
(AdS) \cite{dl} geometries and their GEMS.  There, 
constantly accelerated observers were mapped into similar 
ones in the GEMS.  The resulting Unruh temperatures 
associated with these Rindler motions agreed with those in
the original dS and AdS spaces.  [Actually, AdS has no
real horizon, but temperature is well-defined for 
sufficiently large accelerations and the two methods 
agree  both as to the range where $T$ exists and
to its magnitude.]
In the present paper\footnote{A brief summary of part
of this work was given in \cite{dl2}.}, we will show that the GEMS
approach indeed provides a unified derivation of
temperature for a wide variety of curved spaces,
including general rotating BTZ, Schwarzschild together
with its dS and AdS extensions, and 
Reissner--Nordstrom.  
In each case the usual black-hole (BH) detectors
are mapped into Rindler observers with the correct
temperature as determined from their (constant)
accelerations.  Conversely, we will also  connect surface 
gravity and Unruh temperatures, for  both Rindler 
observers in flat  space and   various
accelerated observers in de-Sitter (dS) and anti de-Sitter 
(AdS) spaces, thereby establishing the equivalence principle 
between constant acceleration and ``true" gravity effects.
We will also consider the associated extensive quantity,
the entropy, and again show the mapping correctly matches
the area of  the GEMS Rindler motion and 
``true" horizons, thereby
confirming the equivalence for entropy as well.

We will first review how  temperature measured by
an accelerated detector in   dS/AdS  geometries, say in D=4, 
is just its Unruh temperature (i.e., Rindler acceleration divided by $2\pi$)
in the D=5 GEMS, by relating 
the corresponding 4- and 5-accelerations.
In this connection we will also explicitly relate surface 
gravity to the associated temperatures. Next we shall treat 
rotating and non-rotating D=3 BTZ spaces
\cite{btz}, \cite{btzh}.  Since BTZ is obtained from AdS 
through geodesic identification, 
we will show that we can use the treatment of Unruh observers 
in AdS to calculate the BH temperature here as well, 
in agreement with earlier results. 
Our final applications will be to Schwarzschild, 
Schwarzschild-dS, Schwarzschild-AdS and Reissner-Nordstrom
spacetimes, where the same connections are made, this time 
the required GEMS extensions having D$\geq 6$.
More generally, it will be seen that  for any geometry 
admitting a group of constantly accelerated
observers which encounter a horizon as they follow a  
``bifurcate" timelike Killing vector field,
 the temperature measured by each observer
is simply $2\pi T=a_{\mbox{}_G}$ when $a_{\mbox{}_G}$ is their acceleration 
as mapped 
into the GEMS.  Finally, we will  establish equivalence of
entropies using the Unruh definition in terms of the 
"transverse" Rindler area \cite{lafl}, together with the fact that horizons map
into horizons. 

%\vspace{.2in}
\mbox{}\\
{\bf 2. Surface gravity-Unruh effect connection in 
dS/AdS}

We begin with a brief summary  of the GEMS approach to 
temperature given in
\cite{dl}, for dS/AdS spaces of cosmological constant  
$\Lambda\equiv\pm 3R^{-2}$;
these are  hyperboloids  in the $D=5$  GEMS $ds^2=
\eta_{\mbox{}_{AB}}(dz^{\mbox{}^A})^{2}(dz^{\mbox{}^B})^{2}$,
\be
\eta_{\mbox{}_{AB}}(z^{\mbox{}^A})^{2}(z^{\mbox{}^B})^{2}=\mp R^{2}.
\ee
Here $A,B=0...4\; ,\; \eta_{\mbox{}_{AB}}=\rm{diag}
(1,-1,-1,-1,\mp 1)$; throughout,
upper/lower signs refer to  dS/AdS respectively.
 We specifically consider $z^{1}=z^{2}=0$
and $z^{4}=Z=\rm{const}$ trajectories, obeying 
 $(z^{1})^2-(z^{0})^{2}=\pm R^{2}\mp Z^{2}\equiv a_{5}^{-2}$. 
Now  the Unruh effect states that flat space detectors  
with constant acceleration $a$  along the 
$x$ direction, whose motions are thus on
$ x^{2}-t^{2}=a^{-2}$, measure temperature $2\pi T=a$. Since our 
embedding space detectors follow precisely such  trajectories 
{\it i.e.}, have a Rindler-like motion with constant acceleration 
$a_{5}$, they measure  
\be
2\pi T=a_{5}=(\pm R^{2}\mp Z^{2})^{-1/2}\equiv (\pm R^{-2}
+a^{2})^{1/2}.
\ee
The last equality expresses the temperature  in term of the D=4 
quantities, using $a_{5}^2=\pm R^{-2}+a^{2}$.

The relation  between Hawking-Bekenstein 
horizon surface gravity $k_{\mbox{}_H}$ 
and the BH temperature (originally found for Schwarzschild BH)
\cite{h1}, \cite{bek} 
\be
T=\frac{1}{2\pi}\frac{k_{\mbox{}_H}}{\sqrt{g_{00}}}\; ,
\ee
where $x^{0}$ is the time-like Killing vector of a detector 
in its rest frame, holds also  for Schwarzschild-AdS and BTZ 
spacetimes \cite{brown}.
 For these latter two, the local
temperature vanishes at infinity, and no Hawking particles are 
present far from the BH: created at the horizon, they do not 
have enough energy to escape to infinity (where the 
``effective potential" becomes infinite). The connection
 (3) between temperature and surface gravity also holds 
\cite{wald} for Rindler motions, reinforcing
the connection between the Hawking and Unruh effects as 
being based on the existence of horizons, whether ``real" 
or just seen by accelerated observers.
In both cases, inserting  the horizon surface gravity
in (3) will give the temperature. To calculate $T$, it is  
convenient to use the detector rest 
frame.\footnote{The vacuum states in these 
timelike Killing coordinate systems are 
Schwarzschild-like. Therefore,
determining the temperature by the (lowest order) 
transition rate obtained from the Wightman function for 
these vacua  gives zero temperature, while the
same calculation for Hawking-Hartle and  Kruskal-like 
vacua gives the temperature (3).} 
The simplest example is the flat space Rindler observer,
best described by
 Rindler coordinates ($\tau , \zeta$) 
\be
ds^2=L^{2}\exp (2\zeta)(d\tau^2-d\zeta^2)-(dy^{2}+dz^{2}).  
\ee
A $\zeta=$const detector (following the time-like Killing 
vector $\xi=\partial_{\tau}$ )
has a constant acceleration
$a=L^{-1}\exp (-\zeta)$. This group of accelerated observers 
sees an event horizon at $\zeta=-\infty$.
Since $\xi$ is perpendicular to the horizon  
(and therefore null)
we can calculate the surface gravity using its
definition \cite{wald}
\be
k^2_{\mbox{}_H}=-\frac{1}{2}(\nabla^{\mu}\xi^{\nu})
(\nabla_{\mu}\xi_{\nu})
\ee
where the right side is to be  evaluated at the horizon. For us
\be
k^2_{\mbox{}_H}=k^2(\zeta=-\infty)=1.
\ee
Inserting $k_{\mbox{}_H}$ in (3) gives the desired result   
\be
2\pi T=L^{-1}\exp(-\zeta)=a.
\ee

Let us show that use of  surface gravity to calculate  
temperature also works for dS/AdS.
Consider first dS with its real horizon, expressed in 
the static coordinates ($t,r,\theta,\phi$)
related to the $z^{A}$  according 
to\footnote{Although this coordinate transformation 
covers only part of the space, it is easy to extend it continuously
to the whole dS, resulting in a global embedding.},
\be
z^{0}\!=\!\sqrt{R^{2}-r^{2}}\sinh(t/R)\;\;\; 
z^{1}\!=\!\sqrt{R^{2}-r^{2}}\cosh(t/R) \;\;\;
\ee
\[
z^{2}\!=\!r\sin\theta\cos\phi\;\;\;z^{3}
\!=\!r\sin\theta\sin\phi\;\;\;
z^{4}\!=\!r\cos\theta .\nonumber
\]
The metric
\be
ds^2=\left[1-\frac{r^{2}}{R^{2}}\right] dt^{2}-
\left[ 1-\frac{r^{2}}{R^{2}}
\right]^{-1}dr^{2}-r^{2}
(d\theta^{2}+\sin^2\theta d\phi^2 )
\ee   
has an intrinsic horizon at $r=R$. It is seen by 
``static" detectors ($r, \theta, \phi$ const), 
or equivalently (choosing $\theta=0$, as is 
allowed by symmetry)
$z^{1}=z^{2}=0$ and $z^{4}=r=Z=$const. They follow the 
time-like Killing vector $\partial_{t}$ and 
have constant acceleration $a=r/(R\sqrt{R^{2}-r^{2}})$. 
Hence by (5), we have
\be
k_{\mbox{}_H}=1/R
\ee
and the temperature measured by these detectors  agrees 
with the known results
of \cite{n},
\be
T=\frac{1}{2\pi} \frac{1}{\sqrt{R^{2}-r^{2}}} 
=\frac{1}{2\pi}
\sqrt{\frac{1}{R^{2}}+a^{2}}.
\ee
In AdS, 
\be
ds^2=\left[1+\frac{r^{2}}{R^{2}}\right] dt^{2}-
\left[ 1+\frac{r^{2}}{R^{2}}
\right]^{-1}dr^{2}-r^{2}
(d\theta^{2}+\sin^2\theta d\phi^2 ),
\ee   
there is no intrinsic horizon. So although $r=\rm{const}$ 
detectors have  constant
acceleration  $a=r/(R\sqrt{R^{2}+r^{2}})<R^{-1}$, 
they will not measure
any temperature. The intrinsic horizon of dS causes 
even inertial
detectors to measure temperature, while in AdS the 
absence of a real horizon causes sufficiently 
slowly ($a<R^{-1}$) accelerated  detectors  not to 
measure one.
There is no contradiction with the Unruh picture: 
as we will see, the GEMS
acceleration $a_{5}^{2}$ becomes negative for 
them\footnote{If we  take the imaginary point
$r_{\mbox{}_H}=\pm iR$  to define the AdS ``horizon" and 
 calculate the surface gravity
at that point,   
 (3) will give, as expected, an imaginary temperature 
$2\pi T=\pm i (R^{2}+r^{2})^{-1/2}= \sqrt{-R^{-2}+a^{2}}$, 
but (by the last equality) the correct temperature formula  for AdS 
\cite{dl}.}$\mbox{}^{,}$\footnote{It 
is also possible to get the AdS result from that of
Schwarzschild-AdS \cite{brown}, not by taking the 
limit $m\rightarrow 0$ but only by setting $m=0$ 
initially. This is exactly like the 
impossibility of reaching  flat 
space by taking the $m\rightarrow 0$ limit of the Hawking 
temperature formula for Schwarzschild.}.
Indeed the ``GEMS temperature" was obtained only for 
$(z^{4})^2={\em const}^2>R^2$ ($a>R^{-1}$) trajectories there 
\cite{dl}. Using the formula 
for time-like trajectories with $a<R^{-1}$ (not  
$(z^{4})^{2}>R^2$ trajectories, but for example the 
$z^{1}=$ const, or the $r=$const case we 
discussed above)  would lead to imaginary $T$: the 
detector will not  measure 
any temperature because it sees no  event horizon, 
hence no loss of information. To calculate the temperature 
using (3) when $a>R^{-1}$ 
it is convenient to use a new coordinate system 
(the one in \cite{dl} is not suitable here since its 
$x^{0}$ is not the time-like Killing vector followed by 
the observers). Instead we introduce an ``accelerated" 
coordinate system  obtained by the GEMS
coordinates defined from the D=4 covering of AdS 
\be
ds^2=\frac{- R^2 +\rho^2}{R^2} d\eta^2-
\frac{R^2}{-R^2 +\rho^2} d\rho^2- \rho^2(d\psi^2+\sinh^2 \psi
d\theta^2)
\ee
as follows:
\be
z^{0}\!=\!\sqrt{-R^{2}+ \rho^{2}}\sinh (\eta/R); ,\;
z^{1}\!=\!\sqrt{- R^{2}+ \rho^{2}}\cosh (\eta/R)
\ee
\[
z^{2}\!=\!\rho\sinh \psi\cos \theta\; ,\;
z^{3}\!=\!\rho\sinh \psi\sin \theta\; ,\;
z^{4}=\rho\cosh\psi.
\]
Here $-\infty<\eta,\psi<\infty, \; -\pi<\theta<\pi$; while this 
coordinate patch only covers the region  $\rho>R$,  it
can be  extended to the entire space. 
Since we are interested in $z^{1}=z^{2}=0 \;\; 
z^{4}=\rm{const}$ trajectories,
 $\psi$ is set to zero, and $\rho$ to a constant $Z$; their accelerations are
$a^2=Z^2R^{-2}(Z^2-R^2)^{-1}>R^{-2}$. For AdS, the horizon appears in this 
``accelerated" frame 
exactly as it did upon transforming from Minkowski to 
Rindler coordinates in flat space.
These trajectories follow the time-like Killing vector 
field $\partial_{\eta}$
which is null at the event horizon $\rho=R$, so (5) gives
\be
k_{\mbox{}_H}=R^{-1}.
\ee
The corresponding temperature, from (3), is
\be
2\pi T=(-R^2 + Z^2)^{-1/2}=(-R^{-2}+a^2)^{1/2}
\ee
which is exactly the result obtained using the kinematical 
behavior of these trajectories in the GEMS,  as well as  
by calculating the transition rate in the 
``non-accelerated" coordinate system.

\mbox{}\\
{\bf 3. BTZ spaces}

In the previous section, we demonstrated the feasibility 
of using  surface gravity (or equivalently 
Hawking--Bekenstein temperature) to calculate the 
temperature measured in dS/AdS, in agreement with that 
obtained by purely kinematical Unruh considerations. 
This immediately raises the converse question: 
calculate Hawking temperature entirely
from  GEMS kinematics when ``real", mass-related, horizons are present.
The simplest candidate for this would seem to be the BTZ black hole solution, 
due to its relation to AdS; we now use our method  to calculate BTZ 
temperature, at least for some observers, and compare with previous
calculations using surface gravity \cite{btz},\cite{carlip}.
            
The general rotating BTZ black hole is described by the 3-metric
\be
ds^2=N^2dt^2-N^{-2}dr^{2}-r^{2}(d\phi+N^{\phi}dt)^2
\ee
\[N^{2}\equiv (r^2-r_{+}^2)(r^2-r_{-}^2)/(r^2 R^2)\;\;\; , \;\;\;
N^{\phi}\equiv -r_{+}r_{-}/(r^2 R).\]
It arises from  AdS upon making the geodesic
identification  $\phi=\phi+2\pi$.
The coordinate transformations to the (2+2) AdS GEMS 
$ ds^{2}=(dz^{0})^2-(dz^{1})^2 -(dz^{2})^2+(dz^{3})^{2}$ are
for, $r\geq r_{+}$ (the extension to $r<r_{+}$ is given in
\cite{btzh})
\be
z^{o}\!=\!R\sqrt{\frac{r^{2}-r_{+}^{2}}{r^{2}_{+}
-r^{2}_{-}}}
\sinh \!\left( 
\frac{r_{+}}{R^2}t\! -\! \frac{r_{-}}{R}\phi \right) \;\;
z^{1}\!=\!R\sqrt{\frac{r^{2}-r_{+}^{2}}{r^2_{+}-r^2_{-}}}
\cosh \!\left( 
\frac{r_{+}}{R^2}t\!-\!\frac{r_{-}}{R}\phi \right)
\ee
\[
z^{2}\!=\!R\sqrt{\frac{r^{2}-r_{-}^{2}}{r^{2}_{+}
-r^{2}_{-}}}\sinh \left( 
\frac{r_{+}}{R}\phi\!-\!\frac{r_{-}}{R^2}t \right)
\;\;\;
z^{3}\!=\!R\sqrt{\frac{r^{2}-r_{-}^{2}}{r^{2}_{+}
-r^{2}_{-}}}\cosh \left( 
\frac{r_{+}}{R}\phi\!-\!\frac{r_{-}}{R^2}t \right) ,
\]
where the constants
$(r_{+},r_{-})$ are related to the mass and angular 
momentum.    This AdS GEMS can serve as the BTZ embedding space for our
purpose. In spite of the fact that there is no longer a one to one mapping 
between it and the BTZ space due to the $\phi$ identification,
following a detector motion with certain initial condition such as $\phi(t=0)=0$
still gives a unique trajectory in the embedding space which is the basic
requirement of our approach  based on the observer's kinematical behavior in
the GEMS: If the detector trajectory maps (without  ambiguity) into an
Unruh one in the GEMS, then we can use it for temperature calculation. 
  
Consider first non-rotating BTZ ($r_{-}=0$)
and  focus on ``static" detectors ($\phi , r$ $=$ const).
These detectors have  constant 3-acceleration
$a=rR^{-1}(r^2-r_{+}^{2})^{-1/2}$, and are
describe by a (fixed)  point in the ($z^{2},z^{3}$) plane (for example 
$\phi=0$ gives $z^{2}=0\;\; z^{3}=$const), and   
  constant accelerated motion in 
($z^{0},z^{1}$) with $a_{4}=r_{+}R^{-1}(r^2-r_{+}^2)^{-1/2}$.
So in the GEMS we have a constant Rindler-like accelerated
motion and the temperature measured by the detector is
\be
2\pi T= a_{4}=r_{+}R^{-1}(r^2-r_{+}^{2} )^{-1/2}=
(- R^{-2}+a^{2} )^{1/2} 
\ee
which is that obtained  using (3), and agrees
with the temperature given by the response function of 
particle detectors
\cite{ortiz}. In the asymptotic limit $r \rightarrow 
\infty$, BTZ tends to AdS,
the acceleration  $a\rightarrow R^{-1}$, which is 
of course the
acceleration of a ``static" detector at infinity in AdS; 
both detectors measure zero temperature\footnote{BTZ formally becomes AdS 
in our coordinates by setting $r_{-}=0$ and
$r_{+}=\pm iR$; (17) and the D=3 version of (12) are the same. This shows 
again  that AdS has a hidden imaginary horizon which causes the threshold in 
the temperature (acceleration smaller than 
$R^{-1}$ measures no temperature).}(no Hawking particle at infinity).
The rotating case is  more complicated.
The Hawking temperature $2\pi T= (Rr)(r^2-r_{+}^2)
^{-1/2}(r^2-r_{-}^2)^{-1/2}
k_{\mbox{}_H}$ ,  $k_{\mbox{}_H}=(r_{+}^2-r_{-}^2)(r_{+}R^2)^{-1}$ 
was calculated \cite{carlip}, \cite{brown} for 
trajectories that follow the time-like Killing vector 
$\xi=\partial_{t}-N^{\phi}\partial_{\phi}$,
i.e. observers that obey $\phi=-N^{\phi}t\;\; r=$const (and hence are 
``static" at infinity). Although they have a constant D=3 acceleration,
$a=(r^4-r_{-}^2 r_{+}^2) / (r^2 R \sqrt{ (r^2-r_{-}^2)(r^2-r_{+}^2)} ) $,
these trajectories do not describe  pure Rindler motion 
in the  GEMS, combining 
accelerated motion in the ($z^{0},z^{1}$) plane 
with a space-like motion in  ($z^{3},z^{2})$. 
Therefore, we cannot use their kinematical behavior in this GEMS to calculate
 the temperature they measure. Exactly the same problem would arise for any AdS
detector with $\psi\neq$const in (14). This particular case resembles
 AdS motion with  $\psi=\alpha(r)t$, $\theta=0$.
Our method can be used only for a group of detectors that maps into a group of
pure Unruh observers in the GEMS.
Hence, it is only possible to use it for those 
observers for whom the map of the detector trajectory  into the 
``transverse" embedding space (for BTZ the  $z^{2},
z^{3}$ plane) is time-independent, $\it{i.e.}$, the detector motion  at any time
is described by a fixed point in that plane.  
There is one group of time-like observers obeying
\be
\phi=\frac{r_{-}}{r_{+}R}t\;\;\; , \;\;\; r=\rm{const}
\ee
that does allow us to use the above GEMS
 and hence to  compare the two calculation of $T$. 
These detectors have a constant
acceleration $a\!=\!(r^{2}-r_{-}^2)^{1/2}(r^{2}-r_{+}^2)
^{-1/2}R^{-1}$ in BTZ and a Rindler-like motion in the GEMS with 
acceleration  
$a_{4}=R^{-1}(r^2_{+}-r^2_{-})^{1/2}(r^{2}
-r_{+}^{2})^{-1/2}$ and therefore measure  
\be
2\pi
T=a_4=\frac{1}{R}\sqrt{\frac{r^2_{+}-r^2_{-}}{r^{2}
-r_{+}^{2}}}=\sqrt{-R^{-2}
+a^{2}}.
\ee
On the other hand, inserting $\phi=\frac{r_{-}}
{r_{+}l}t$ into (17) gives
\be
ds^2=\frac{(r^2-r_{+}^2)(r_{+}^2-r_{-}^2)}
{r_{+}^2 R^2}dt^{2}
-\frac{r^2 R^2}{(r^2-r_{+}^2)(r^2-r_{-}^2)} dr^2
\ee
which show us that they follow the time-like 
Killing vector field $\xi=\partial_{t}$ for this
metric (or $\xi=\partial_{t}+r_{-}(r_{+}R)^{-1}
\partial_{\phi}$ if we use (17)) and see an event 
horizon at the metric's own ``real" event horizon 
$r=r_{+}$. The surface gravity is
\be
k_{\mbox{}_H}=(r_{+}^2-r_{-}^2)/(r_{+}R^{2}),
\ee
which is the same as that calculated for the other group by using the other 
Killing vector. This equivalence exists since both have the 
same horizon $r=r_{+}$
and the Killing vectors they follow are the 
same there. Any scaling problems are avoided since 
we used a common coordinate system.
 While  surface gravity can be obtained  
from either of the metrics (22) or (17),
the appropriate  $g_{00}$ must taken from  (22)
because only there is $x^{0}$ the time-like Killing 
vector followed by the observer.  This gives
\be
2\pi T=\frac{1}{R}\sqrt{\frac{r^2_{+}
-r^2_{-}}{r^{2}-r_{+}^{2}}},
\ee
exactly the  result  obtained by using the GEMS. 
Finally, we note that a common alternative definition 
of BH temperature is 
to scale $T$ by $\sqrt{g_{00}}$: 
$T_{0}=\sqrt{g_{00}}T=k_{\mbox{}_H}/2\pi$; as distinct
from the local temperature $T$, it is $T_{0}$ 
that enters into the BH thermodynamics relations.
Since there is one observer (the $r=r_{+}$ one) that belongs to both of the
different observer groups ($\phi=-N^{\phi}t$ and $\phi=
r_{-}t/(r_{+}R)$), and since $T_{0}$ is a global feature of all the members in
the group, it is obvious that both groups should give the same 
temperature (this of course could be seen immediately from their surface gravity
equivalence). On the other hand, it should be  no surprise that detectors in 
the two different observer groups measure different temperatures even though
 their absolute accelerations are  the same 
(the Rindler relation $2\pi T=a_4$ does not apply
to the $\phi=-N^{\phi}t$ group) because the temperature
 $T$ is observer-dependent in general. 
Since BTZ is asymptotically AdS, both detectors will again measure zero 
temperature at $r \rightarrow \infty$, where $a\rightarrow R^{-1}$.

\mbox{}\\
{\bf 4. Schwarzschild and related geometries.}

We now come to spaces with ``more manifest" real horizons.
Once a GEMS has been found (they always exist \cite{goenner}) for the desired 
physical space, it is a mechanical procedure, using the familiar embedding 
Gauss-Codazzi-Ricci equations to relate 
constant acceleration $a_{\mbox{}_G}$ in GEMS to the embedded 
space physics; this is also possible when (as for 
Schwarzschild) the GEMS is more than one dimension 
higher.
The acceleration of detectors that follow a time-like 
Killing vector $\xi$ in the
physical space is \cite{n} $a=\nabla_{\xi} \xi/|\xi|^2$ where 
$|\xi|$ is the norm of $\xi$.
It is related to $a_{\mbox{}_G}$ in the GEMS according to
\be
a_{\mbox{}_G}^2=a^2+\alpha^{2}|\xi|^{-4}     
\ee
where $\alpha$ is the second fundamental form
\cite{goenner}.  Thus the temperature should simply be 
$2\pi T= a_{\mbox{}_G}=[a^2+\alpha^{2}|\xi|^{-4}]^{1/2}$.
One should not, however, assume from this formula that 
there is always a    
temperature, since in fact $\alpha^2$ 
need not always be positive (it is
$\alpha^2 |\xi |^{-4}=-R^{-2}$ in AdS). After all, it is only when 
$a_{\mbox{}_G}^2$ 
is non-negative that the Unruh description itself is 
meaningful in a flat space.

We apply these 
ideas first to the three types of Schwarzschild (vacuum) spaces, beginning with
the usual case without cosmological constant; it can be globally embedded 
in flat D=6 
\be
ds^{2}=(dz^{0})^2-(dz^{1})^2-(dz^{2})^2-(dz^{3})
^2-(dz^{4})^2-(dz^{5})^2 ,
\ee
using the coordinate transformation \cite{fronsdal}, 
\[
z^{0}= 4m\sqrt{1-2m/r}\sinh (t/4m)\;\;\;\;z^{1}= 
4m\sqrt{1-2m/r}\cosh (t/4m)
\]
\be
z^{2}=\int dr \sqrt{ (2mr^2+4m^2r+8m^{3})/r^{3}}
\ee
\[
z^{3}=r\sin \theta\sin\phi  \;\;\;\;
z^{4}=r\sin\theta\cos\phi   \;\;\;\;
z^{5}=r\cos\phi.  \]
This transformation can be extended to cover the $r<2m$ interior thanks to the
analyticity of $z^2(r)$ in $r>0$. Indeed, the extension is just the maximal
Kruskal one \cite{kruskal}.
The original Hawking detectors (moving according to
constant $r$, $\theta$, $\phi$), are here 
Unruh detectors;  their six-space 
motions are the now familiar hyperbolic trajectories
\be
(z^{1})^2-(z^{0})^2=16m^2(1-2m/r)=a_{6}^{-2}.
\ee
Hence we immediately infer the  local  Hawking and 
BH temperatures
\be
 T= a_{6}/2\pi=(8\pi m\sqrt{1-2m/r})^{-1}
\;\; ,\;\; T_{0}=\sqrt{g_{00}}T=
(8\pi m)^{-1}.
\ee 
It should be cautioned that use of incomplete 
embedding
spaces, that cover only  $r>2m$ (as for example 
in \cite{rosen}), 
will lead to  observers there for whom
there is no event horizon, no loss of information, 
and no temperature.

The above calculation is easily generalized to 
Schwarzschild-AdS spaces (where $1-2m/r$ is replaced by $1-2m/r+r^2/R^2$) 
using a D=7 GEMS with an additional timelike dimension $z^6$,
\[ z^{0}\!=\!k_{\mbox{}_H}^{-1}\sqrt{1\!-\!2m/r\!+\! 
r^2/R^2}\sinh (k_{\mbox{}_H}t)\; , \;\;
   z^{1}\!=\!k_{\mbox{}_H}^{-1}\sqrt{1\!-\!2m/r\!+\!
r^2/R^2}\cosh (k_{\mbox{}_H}t)
\]
\be
z^{2}=\int\frac{R^3+Rr^2_{\mbox{}_H}}{R^2+3r^2_{\mbox{}_H}}
\sqrt{\frac{r^2r_{\mbox{}_H}+rr^2_{\mbox{}_H}+r^{3}_{\mbox{}_H}}{r^3(r^2
+rr_{\mbox{}_H}+r^2_{\mbox{}_H}+R^2)}}dr
\ee
\[ 
z^{6}=\int\sqrt{\frac{(R^4+10R^2r_{\mbox{}_H}^2+9r^4_{\mbox{}_H})
(r^2+rr_{\mbox{}_H}+r^2_{\mbox{}_H})}{r^2+rr_{\mbox{}_H}+r^2_{\mbox{}_H}
+R^2}}\frac{dr}{R^2+3r_{\mbox{}_H}^2}
\] 
and ($z^{3},\; z^{4},\; z^{5}$) as in (27); 
$k_{\mbox{}_H}\!=\!(R^2+3r_{\mbox{}_H}^2)/2r_{\mbox{}_H}R^2$ is the surface 
gravity at the root 
$r_{\mbox{}_H}$ of  $(1-2m/r+r^2/R^2)\!=\!0$.
Using this GEMS\footnote{It is easy to see that when 
$R\rightarrow\infty$ (the
Schwarzschild limit), $r_{\mbox{}_H}\rightarrow 2m$, $k_{\mbox{}_H}\rightarrow 
(2r_{\mbox{}_H})^{-1}$ and 
$z^2$ becomes identical to the Schwarzschild one while $z^6$ vanishes, 
so that we indeed get back the Schwarzschild GEMS. 
When $m=0$ (the AdS limit), we have 
$r_{\mbox{}_H}^2=k_{\mbox{}_H}^{-2}=-R^2$
 and both $z^2$ and $z^6$ vanish, leaving the AdS GEMS of (12).},
we obtain  $2\pi T=k_{\mbox{}_H}(1-2m/r+ r^2 /R^{2})^{-1/2}$, 
 equal to that  calculated in \cite{brown}. 
[It may seem that we 
have the freedom to choose an arbitrary  constant 
rather than $k_{\mbox{}_H}$ in $z^0$ and $z^1$ and thereby get a different 
temperature. But for any other choice,  $z^2$ and
$z^6$ cannot be chosen so that both their integrands are finite at the 
horizon.
Hence, such  embedding spaces are not 
global, cover only the area outside the horizon and cannot be extended; they
are therefore excluded.]     

For Schwarzschild-dS, which differs formally from Schwarzschild-AdS by 
$R^2\rightarrow 
-R^2$, there are two real
horizons ($r_{+},r_{-}$) in general, both of which could be
seen by physical detectors (such as constant $r$, 
with $r_{-}<r<r_{+}$). This
requires use of a GEMS that captures both horizons. 
Although we have not tried to define this bigger GEMS, 
we do reproduce the known results \cite{gibbons} for temperature of
each separate horizon, by using (30), 
with $R^2\rightarrow
-R^2$ and the respective  $k_{\mbox{}_H}(r_{+})$, $k_{\mbox{}_H}(r_{-})$.
[Our method becomes meaningless for  the extremal 
$(r_{+}=r_{-})$ case since the whole Rindler wedge vanishes there.] 

We turn now to an example with matter, the 
Reissner--Nordstrom solution with
\be
ds^{2}=\left( 1-\frac{2m}{r}+\frac{e^{2}}
{r^{2}}\right) dt^{2}
-\left( 1-\frac{2m}{r}+\frac{e^{2}}{r^{2}}\right)^{-1}dr^2
-r^{2}(d\theta^2+\sin^{2}\theta d\phi^{2}).
\ee
Although there are two horizons ($r_{\pm}=m\pm
\sqrt{m^{2}-e^{2}}$) in the 
nonextremal case ($m>e$),
it is still simple to calculate the temperature via the 
embedding space.
 As explained earlier, a reliable GEMS has to cover (or be extendable to cover)
both sides of the horizon,  or else 
there is no loss of information for a detector in 
that space. But  physical ($r>r_{+}$) $r=$const 
Reissner--Nordstrom detectors 
are aware only of the existence of one horizon $r_{+}$, 
unlike the physical
Schwarzschild-dS  $r$=const detectors 
($r_{+}>r>r_{-}$) that see two
horizons. Therefore, it is enough to use as the embedding space, again with
an added timelike $z^6$ dimension,
\[
z^{0}\!=\!k_{\mbox{}_H}^{-1}\sqrt{1\!-\!2m/r\!+\! e^2/r^2}
\sinh (k_{\mbox{}_H}t)\; , \;\;
  z^{1}\!=\!k_{\mbox{}_H}^{-1}\sqrt{1\!-\!2m/r\!+\! e^2/r^2}
\cosh (k_{\mbox{}_H}t)\; 
\]
\be 
z^{2}=\int \left(\frac{r^2(r_{+}+r_{-})+r_{+}^2(r+r_{+})}{r^{2}(r-r_{-})}
\right) ^{1/2} dr
\ee
\[
z^6=\int\left(\frac{4r_{+}^{5}r_{-}}{r^{4}(r_{+}-r_{-})^2}\right)^{1/2} dr
\]
 with ($z^{3}$, $z^{4}$, $z^{5}$) as in (27), 
and $k_{\mbox{}_H}=k(r_{+})=(r_{+}-r_{-})/2r^2_{+}$.     
[In the neutral, $e=0$,
limit, $z^{6}$ vanishes and this GEMS becomes the (D=6) Schwarzschild one.]
Even though it does not reach down to
$ r \leq r_{-} $, this embedding suffices, because it covers $r_{+}$, for 
the purpose of calculating the Reissner--Nordstrom temperature in the
nonextremal case\footnote{To be sure, our mapping approach has limitations: 
since the Rindler horizon of
the GEMS is  Killing bifurcate, one can only map from spaces whose 
horizons also are;
this excludes the strictly extremal ($m=e$) Reissner-Nordstrom, which is also
exceptional from the D=4 point of view \cite{ted}.}.
 It is clear from (32) that the 
relevant D=7 acceleration  
$a_{7}=((z^1)^2-(z^0)^2)^{-1/2}=(r_{+}-r_{-})/(2r^2_{+}\sqrt{1-2m/r+
e^2/r^2})$  gives the correct Hawking temperature
$T=(r_{+}-r_{-})/(4\pi r_{+}^{2}\sqrt{1-2m/r+ e^2/r^2})$.

\mbox{}\\
{\bf 5. Entropy}

We turn now to the ``extensive" companion of temperature,
the entropy.  For those of our curved spaces 
with intrinsic horizons, and at our semiclassical 
level, entropy is just one quarter of the horizon 
area.  Entropy can also be defined for a Rindler wedge
\cite{lafl}, using arguments similar to those used
originally \cite{g2} for Schwarzschild and dS.  
Here the relevant
area is that of the null surface $x^2 - t^2 = 0$.
This ``transverse" area is in general infinite
for otherwise unrestricted Rindler motion, being
just the cartesian $\int dy\, dz$ for D=4, say.
For our purposes, however, we must evaluate this
area subject to the embedding constraints, and we
shall see, the resulting integral becomes finite
and agrees with that of the original horizon.  [This
is not a tautology: we are not initially writing the
original horizon area in embedding coordinates,
although the result is indeed that real and embedding
horizon areas agree. Nor is it a surprise: we have insured that (when present)
horizons map to horizons.]

Let us begin with the dS case, where the Rindler
horizon condition is
$(z^1)^2 - (z^0)^2 = 0$ which was $Z=R$, and of course
$(z^2)^2 + (z^3)^2 + (z^4)^2 = R^2$.  Thus the integration
over $dz^2 \, d z^3 \, dz^4$ is restricted to
the surface of the sphere of radius $R$, precisely
that of the true horizon.  The AdS case differs,
(as expected from lack of an intrinsic horizon) and
the corresponding restrictions are
$(z^1)^2 - (z^0)^2 = 0$ which again implies $Z=R$, but now
$(z^2)^2 + (z^3)^2 - (z^4)^2 = -R^2$, and the area of this
hyperboloidal surface diverges, having no further
restrictions.
For comparison with the BTZ case below, the cause of
the infinity can be traced to the fact that the
limits on the $z^4$ integral are $\pm R \sinh
\psi$, with $-\infty < \psi < \infty$.

We now see how the BTZ solution 
leads to a finite Unruh area due to the periodic 
identification of $\phi$ mod $2\pi$.  The
$(z^1)^2 - (z^0)^2 = 0$ Rindler horizon condition
implies $r=r_+$, while 
$(z^3)^2 - (z^2)^2 = R^2 (r^2-r^2_-)/(r^2_+-r^2_-)=R^2$
still looks hyperbolic. However, the relevant 
bounds on $z^3$ due to the periodicity are $R\sinh (r_{+}\pi/R^2)$
and $-R\sinh (r_{+}\pi/R^2)$ for the nonrotating case, so that one has the 
integral
\[
\int^{R \sinh (r_+\pi/R)}_{-R \sinh (r_+\pi/R)}
\int^{\sqrt{R^2+(z^2)^2}}_{0} \delta(\sqrt{(z^3)^2-(z^2)^2}-R) dz^3dz^2 =
\int^{R \sinh (r_+\pi/R)}_{-R \sinh (r_+\pi/R)}
\frac{R}{\sqrt{R^2+(z^2)^2}} dz^2,
\] 
(and limits $R \sinh (r_+\pi/R - r_-t/R^2)$ and
$R \sinh (r_+\pi /R + r_-t/R^2)$ in the rotating case)
and it yields
the desired area integral $2\pi r_+$.  
It is clear that the limits differ from the AdS ones
precisely in having the ``angle's" bounds be finite
here.

The Schwarzschild case, where there are two additional
dimensions in the transverse area,
$\int dz^2 \ldots dz^5$, is correspondingly subject to
three constraints:  $(z^1)^2-(z^0)^2=0$ leads to $r=2m$,
(horizon to horizon mapping) $z^2=f(r)$ and $(z^3)^2 + (z^4)^2 + (z^5)^2 = r^2$.
Thus the $z^2$ integral, $\int dz^2 \d (z^2-f(r))$,
is unity, while the remaining integrals of course
reproduce the area of the $r=2m$ sphere in D=3.
The Reissner--Nordstrom and the Schwarzschild-AdS calculations are essentially
the same\footnote{We may still use the Reissner--Nordstrom partial GEMS for 
the area calculation, since it covers the original $r=r_{+}$ horizon.},
 except that in these cases there are three (rather than two) additional 
dimensions, 
and
four constraints: $(z^1)^2-(z^0)^2=0$ leads to $r=r_{+}$,
$z^2=f_1(r)$, $z^6=f_2(r)$ and $(z^3)^2 + (z^4)^2 + (z^5)^2 = r^2$.
Thus the $z^2,\;z^6$ integrals, $\int dz^2dz^6
\delta(z^2-f_1(r))\delta(z^6-f_2(r))$, are unity, and the $z^3, \;z^4,\;z^5 $
integrals gives the desired area, that of the $r=r_+$ sphere. Having two 
separate horizons, the Schwarzschild-dS system is more
delicate to handle, but just as for temperature, we can calculate entropy for 
each horizon 
separately, to obtain the corresponding D=4 results \cite{gibbons}.

\mbox{}\\                                         
{\bf 6. Summary}

We have formulated a uniform mechanism for reducing 
curved space BH horizon temperatures and entropies 
to those of the kinematical Unruh effect due to Rindler motion 
in their GEMS. The latter must of course first 
be found and cover enough of the underlying
space to include the horizon in question. This 
method has been applied to a
variety of ``true" BH spacetimes, both vacuum ones like BTZ, 
Schwarzschild and its dS/AdS extensions, as well as
 Reissner-Nordstrom.
 It would be interesting 
to consider other possible applications of GEMS, for example to superradiance
in rotating geometries.

We thanks M. Banados for useful discussions of BTZ.
This work was supported by NSF grant PHY-9315811 and  by the Fishbach
Foundation for OL.

\end{document}